\documentclass[11pt]{article}
\usepackage[margin=1in]{geometry}
\usepackage{amsmath,amssymb,amsfonts}
\usepackage{graphicx}
\usepackage{booktabs}
\usepackage{microtype}
\usepackage[hidelinks]{hyperref}
\usepackage{caption}
\usepackage{float}
\usepackage{mathtools}
\usepackage{enumitem}
\newtheorem{proposition}{Proposition}
\newtheorem{corollary}{Corollary}

\title{From Gaussian Fading to Gilbert-Elliott:\\Bridging Physical and Link-Layer Channel Models in Closed Form}
\hypersetup{pdftitle={From Gaussian Fading to Gilbert-Elliott: Bridging Physical and Link-Layer Channel Models in Closed Form}}
\author{Bhaskar Krishnamachari\thanks{\texttt{bkrishna@usc.edu}} \quad Victor Gutierrez\thanks{\texttt{vagutier@usc.edu}}\\[2mm]
Viterbi School of Engineering, University of Southern California}
\date{April 2026}

\begin{document}
\maketitle

\begin{abstract}
Dynamic fading channels are modeled at two fundamentally different levels of abstraction.  At the physical layer, the standard representation is a correlated Gaussian process, such as the dB-domain signal power in log-normal shadow fading.  At the link layer, the dominant abstraction is the Gilbert-Elliott (GE) two-state Markov chain, which compresses the channel into a binary ``decodable or not'' sequence with temporal memory.  Both models are ubiquitous, yet practitioners who need GE parameters from an underlying Gaussian fading model must typically simulate the mapping or invoke continuous-time level-crossing approximations that do not yield discrete-slot transition probabilities in closed form.

This paper provides an exact, closed-form bridge.  By thresholding the Gaussian process at discrete slot boundaries, we derive the GE transition probabilities via Owen's $T$-function for any threshold, reducing to an elementary arcsine identity when the threshold equals the mean.  The formulas depend on the covariance kernel only through the one-step correlation coefficient $\rho = K(D)/K(0)$, making them applicable to any stationary Gaussian fading model.

The bridge reveals how kernel smoothness governs the resulting link-layer dynamics: the GE persistence time grows linearly in the correlation length~$T_c$ for a smooth (squared-exponential) kernel but only as~$\sqrt{T_c}$ for a rough (exponential/Ornstein--Uhlenbeck) kernel.  We further quantify when the first-order GE chain is a faithful approximation of the full binary process and when it is not, reconciling two diagnostics, the one-step Markov gap and the run-length total-variation distance, that can trend in opposite directions.  Monte Carlo simulations validate all theoretical predictions.
\end{abstract}

%% ====================================================================
\section{Introduction}
%% ====================================================================
Two simple, standard mathematical models dominate the description of time-varying wireless channels, each suited to a different level of abstraction.

At the physical layer, channel dynamics are captured by a correlated Gaussian process.  The canonical example is log-normal shadow fading, where the received signal power in~dB follows a Gaussian random process with a specified temporal covariance kernel~\cite{gudmundson1991}; this representation preserves the full amplitude-level evolution and correlation structure of the fading, and coherence time is defined in terms of the kernel's correlation length~\cite{goldsmith2005}.

At the link layer, the same channel is viewed through a coarser lens: each time slot produces either a successful or a failed transmission, and the relevant dynamics are captured by a binary process with temporal memory.  The Gilbert-Elliott (GE) two-state Markov chain~\cite{gilbert1960,elliott1963} is the standard such model, compressing a complicated fading environment into a tractable ``good/bad'' abstraction that retains burstiness and is widely used in ARQ analysis, queueing, and networked control.

These two models are complementary: they describe the same physical channel at different layers of abstraction.  In principle, the link-layer GE model should be \emph{derivable} from the PHY-layer Gaussian model by thresholding the continuous process into binary states.  Yet despite the ubiquity of both models, an exact, closed-form bridge between them has been lacking.  Practitioners who need GE parameters consistent with an underlying Gaussian fading model must either simulate the mapping numerically or invoke continuous-time level-crossing rate (LCR) approximations inherited from classical Rayleigh fading theory~\cite{rice1944}, which yield continuous-time crossing rates rather than discrete-slot Markov transition probabilities.

This paper provides the missing bridge.

\subsection{Contributions}

The specific contributions are:
\begin{enumerate}[leftmargin=*]
\item \textbf{Exact PHY-to-link-layer bridge.}  Closed-form GE transition probabilities that exactly match the one-step conditional statistics of the thresholded Gaussian process, for any stationary covariance kernel (Section~\ref{sec:closedform}).  The expressions reduce to Owen's $T$-function~\cite{owen1956} for general thresholds (Section~\ref{sec:owen}) and to an elementary arcsine identity for the symmetric threshold via Sheppard's classical result~\cite{sheppard1900} (Section~\ref{sec:arcsin}).

\item \textbf{Kernel-dependent scaling laws.}  A formal asymptotic expansion revealing that the GE persistence time $\mathbb{E}[T_{\mathrm{GE}}]$ grows linearly in the correlation length~$T_c$ for a smooth (squared-exponential) kernel but only as~$\sqrt{T_c}$ for a rough (exponential) kernel, showing that PHY-layer kernel smoothness determines the link-layer persistence behavior (Section~\ref{sec:asymptotics}).

\item \textbf{Fidelity analysis of the bridge.}  A dual-kernel study quantifying when the first-order GE approximation is faithful and when it breaks down, reconciling two diagnostics, the Markov gap and the run-length total-variation distance, that can trend in opposite directions, with a mathematical explanation for why (Section~\ref{sec:markov-accuracy}, Section~\ref{sec:numerical}).
\end{enumerate}

\subsection{Paper outline}
Section~\ref{sec:related} reviews related work.  Section~\ref{sec:system} introduces the system model and both covariance kernels.  Section~\ref{sec:closedform} derives the exact cross-layer bridge: GE transition probabilities in closed form via Owen's $T$-function and the arcsin special case.  Section~\ref{sec:markov-accuracy} quantifies the fidelity of the bridge via the Markov gap and run-length diagnostics, and explains why the two measures can trend in opposite directions.  Section~\ref{sec:asymptotics} derives the kernel-dependent asymptotic scaling laws.  Section~\ref{sec:numerical} presents dual-kernel numerical results with Monte Carlo validation and a quantitative summary table.

%% ====================================================================
\section{Related Work}
\label{sec:related}
%% ====================================================================

\subsection{Origins of the Gilbert--Elliott model and finite-state channels}
The two-state channel with memory commonly referred to as the Gilbert--Elliott (GE) model traces back to early burst-noise channel modeling for telephone circuits. Gilbert introduced a two-state finite-state channel model to study capacity under burst noise \cite{gilbert1960}, and Elliott subsequently used closely related two-state machinery to evaluate error rates for codes on burst-noise channels \cite{elliott1963}. These works established the modern interpretation of GE parameters as (i) Markovian state persistence (good $\leftrightarrow$ bad switching rates) and (ii) state-conditional error behavior.

Beyond the original two-state formulation, classical surveys on channels with memory emphasize why even simple Markovian error processes can materially affect coding and ARQ performance, motivating tractable abstractions that retain temporal correlation \cite{kanal_sastry1978}. In wireless communications, GE is often treated as the $2$-state special case of a finite-state Markov channel (FSMC) used to approximate time-correlated fading \cite{sadeghi2008fsmc_survey}.

\subsection{FSMC fading models from thresholding, level-crossing rates, and fade durations}
A dominant methodology for FSMC modeling of wireless fading begins with a continuous-valued fading process (e.g., Rayleigh/Clarke-type models) and partitions an instantaneous channel-quality variable (envelope or SNR) into discrete states. Transition probabilities are then obtained using second-order fading statistics such as the level-crossing rate (LCR) and average fade duration (AFD), which are rooted in classical level-crossing theory \cite{rice1944} and specialized to mobile radio fading models \cite{clarke1968,jakes1974}.

Canonical FSMC constructions for Rayleigh fading include interval-partition and adjacent-transition models that compute transition probabilities consistent with the fading dynamics \cite{wangmoayeri1995fsmc,zhangkassam1999fsmc}. The survey in \cite{sadeghi2008fsmc_survey} systematizes these derivations, emphasizing the accuracy--complexity tradeoff (number of states, partition criteria, memory order) and the role of verification against simulations or measurements.

\paragraph{Positioning relative to this thread.}
In contrast to LCR/AFD-based approximations (which are intrinsically tied to continuous-time crossings and often assume only adjacent transitions), our contribution targets the binary (two-state) case and derives \emph{one-step} GE transition probabilities in closed form directly from the joint law of two adjacent samples of an underlying stationary Gaussian surrogate.  In the continuous-time limit ($D\to 0$), the one-step crossing probability $N/D$ converges to the Rice level-crossing rate~\cite{rice1944}; what this paper adds beyond LCR/AFD is (a)~a discrete-slot transition matrix rather than continuous-time rates, (b)~the elementary arcsin closed form for $S=0$, and (c)~an explicit Markov-gap analysis quantifying when the first-order approximation suffices.

\subsection{Two-state Markov link abstractions in networking, ARQ, queueing, and control}
Two-state Markov abstractions are widely used as link-layer models to capture burstiness in packet errors and losses, enabling tractable protocol analysis. In wireless ARQ analysis, Markov channel models are commonly assumed to quantify throughput and delay under correlated fading and burst errors \cite{zorzi1997arq,kimkrunz1999arq_delay}. In cross-layer queueing and QoS modeling, Markovian channel dynamics interact with finite buffers and adaptive transmission policies, producing analyzable expressions for loss rate, throughput, and delay \cite{liuzhougiannakis2005queue_amc,moltchanov2006crosslayer}.

A second major consumer of GE-style models is networked estimation and control with packet drops. The i.i.d. intermittent-observation model \cite{sinopoli2004intermittent} is a common baseline, while Markovian loss models capture temporally correlated drop processes and lead to refined stability and optimal-policy results \cite{huangdey2007markov_losses,wushiandersonjohansson2017ge_kalman,chakravortymahajan2020remote_est}. In these domains, GE parameters are often treated as fitted-from-data or scenario-chosen quantities.

\paragraph{Positioning relative to this thread.}
Our closed-form one-step mapping provides a principled route to parameterize (or sanity-check) the failure and recovery rates used in ARQ, queueing, and control analysis from an underlying correlation model and sampling interval, and our induced dwell-time metric offers an interpretable proxy for burst length that directly enters many higher-layer performance calculations.

\subsection{Gaussian thresholding and classical arcsine identities}
The transformation of a continuous Gaussian process into a binary sequence via thresholding is mathematically linked to long-standing results on bivariate normal quadrant probabilities and arcsine laws. Early work on normal correlation integrals and subsequent modern references show that, for standard bivariate normal variables with correlation $\rho$, symmetric-threshold (e.g., sign) events admit elementary expressions involving $\arcsin(\rho)$ \cite{sheppard1900}. In statistical signal processing, closely related arcsine relationships appear in the analysis of clipped/quantized Gaussian signals \cite{vanvleck_middleton1966_clipped}, and more general tools for expectations through nonlinearities under Gaussianity include Bussgang's theorem \cite{bussgang1952} and Price's theorem \cite{price1958}. Related early analyses compute correlation functions of nonlinearly limited Gaussian noise \cite{baum1957}.

\paragraph{Positioning relative to this thread.}
While arcsine identities for bivariate normal quadrant probabilities are classical, they are less commonly translated into communications-style link abstractions with explicit GE transition parameters and dwell-time metrics. Our symmetric-threshold reduction specializes the mapping from bivariate Gaussian statistics to GE parameters into an elementary closed form, enabling direct analytical scaling laws between physical correlation parameters and link-layer burstiness.

\subsection{Model mismatch, approximation error, and empirical validation}
A thresholded process induced by an underlying fading process generally exhibits memory beyond order one; hence, a first-order GE chain should be understood as an approximation that matches certain low-order statistics (notably one-step transitions) rather than as an exact generative mechanism. Several works explicitly test or quantify the adequacy of first-order Markov assumptions in fading/FSMC settings \cite{wangchang1996verify,tanbeaulieu2000firstorder,zorzi1995accuracy}. Government and technical reports also examine Markov character questions for Rayleigh fading and related models \cite{dalke_hufford2005_markov}.

For binary error sequences and FSC modeling, empirical comparisons of GE models with higher-order Markov or hidden Markov alternatives demonstrate when increasing memory order materially improves fit (e.g., via autocorrelation or variational distance metrics) \cite{pimentel2002its_fsc,pimentel2004rician_fsmc}. Trace-driven fitting studies compare GE and higher-order models using multi-timescale second-order statistics and burst-length distributions, illustrating both the usefulness and limitations of simple models \cite{hasslingerhohlfeld2008ge_fit}. Recent measurement-driven studies continue to fit GE parameters to wireless traces (e.g., LTE/Wi-Fi) to study reliability, burst length, and multi-connectivity benefits \cite{nielsen2019multiconnectivity}, while systems work on low-power links highlights persistent temporal dynamics and burstiness phenomena that motivate such abstractions \cite{zunigakrishnamachari2004transitional,srinivasan2008betafactor}.

%% ====================================================================
\section{System model}
\label{sec:system}
%% ====================================================================
In wireless propagation, the well-known log-normal shadow-fading model \cite{gudmundson1991} represents the received signal power in dB as a correlated Gaussian random process. More generally, any fading mechanism whose dB-scale fluctuations are approximately Gaussian and temporally correlated leads to the same mathematical framework. We therefore consider a stationary zero-mean Gaussian process $X(t)$ (representing, e.g., the dB-scale signal power relative to its local mean) with covariance kernel
\begin{equation}
K(t,t') = \sigma^2 \exp\!\left(-\frac{(t-t')^2}{T_c^2}\right),
\label{eq:kernel}
\end{equation}
where $\sigma^2$ is the marginal variance (the shadow-fading spread in dB$^2$) and $T_c$ is the coherence parameter controlling the temporal correlation of the fading process. Larger $T_c$ produces smoother trajectories and slower temporal evolution.

\paragraph{Remark: choice of covariance kernel and generality of the results.}
The squared-exponential (Gaussian) kernel in \eqref{eq:kernel} is a stylized choice; for example, the canonical Clarke/Jakes model for Rayleigh fading leads to a Bessel-function autocorrelation $J_0(2\pi f_D \tau)$.  However, the core derivations in Sections~\ref{sec:closedform}--\ref{sec:arcsin} depend on the covariance kernel \emph{only} through the one-slot correlation coefficient $\rho = K(D)/K(0)$.  Any stationary Gaussian fading model can therefore be plugged in by computing its own $\rho$ at lag~$D$.  The squared-exponential kernel is used here because it (a)~enables the clean asymptotic expansion in Section~\ref{sec:asymptotics}, and (b)~arises naturally in the dB domain of log-normal shadow fading, where spatial/temporal correlations are well modeled by a Gaussian kernel \cite{gudmundson1991}.

To explore the effect of kernel smoothness, we also consider the exponential (Ornstein--Uhlenbeck) kernel
\begin{equation}
K_{\mathrm{exp}}(t,t') = \sigma^2 \exp\!\left(-\frac{|t-t'|}{T_c}\right),
\label{eq:kernel-exp}
\end{equation}
which gives $\rho = e^{-D/T_c}$ and produces a first-order Markov process in continuous time.  Although the underlying OU Gaussian process is Markov, thresholding does \emph{not} preserve this property: knowing both $B(n)$ and $B(n{-}1)$ provides more information about $X(nD)$ than $B(n)$ alone, because a recent state transition constrains $X(nD)$ to lie near the threshold.  Nevertheless, the two kernels produce qualitatively different Markov-gap behavior, making the comparison instructive (Section~\ref{sec:markov-accuracy}).

\paragraph{Remark: $T_c$ versus classical coherence time.}
In Rayleigh/Clarke-Jakes fading, coherence time is conventionally defined via an autocorrelation threshold (e.g., the lag at which $|R(\tau)/R(0)|$ drops below $0.5$) or through the Doppler spread $f_D$ as $T_{\mathrm{coh}} \propto 1/f_D$.  Here, $T_c$ is the \emph{kernel correlation-length parameter}: for the squared-exponential kernel, the one-slot correlation is $\rho = e^{-D^2/T_c^2}$, while for the exponential kernel it is $\rho = e^{-D/T_c}$.  The mapping from $T_c$ to a specific autocorrelation level depends on the kernel shape, but in all cases the single parameter $\rho = K(D)/K(0)$ absorbs the kernel-specific details, so the closed-form GE expressions derived below apply universally once $\rho$ is known.

Now choose a slot duration $D>0$ and a threshold $S$, and sample the process at slot boundaries. Define the binary sequence
\begin{equation}
B(n) =
\begin{cases}
0, & X(nD) < S,\\[1mm]
1, & X(nD) \ge S.
\end{cases}
\label{eq:binary-def}
\end{equation}
The resulting binary process is \emph{not} exactly first-order Markov when the kernel is squared-exponential (see Section~\ref{sec:markov-accuracy} for a detailed analysis), but it can be approximated by a two-state Gilbert-Elliott chain with one-step transition probabilities
\begin{equation}
p_{01} = \Pr(B(n+1)=1\mid B(n)=0),
\qquad
p_{10} = \Pr(B(n+1)=0\mid B(n)=1).
\label{eq:ge-transitions}
\end{equation}

\subsection{Steady-state expected persistence time of the GE chain}
\label{sec:persistence}
For the binary GE abstraction, we define the \emph{expected persistence time} as the expected duration until the next state change when the chain is observed at a random slot boundary in steady state.  This quantity characterizes how long the binary link state is expected to persist, and should be distinguished from the analog-channel coherence parameter~$T_c$, which controls the temporal smoothness of the continuous Gaussian process.

If the current state is $0$, the dwell time until the next transition is geometric with mean $1/p_{01}$ slots. Likewise, if the current state is $1$, the corresponding mean is $1/p_{10}$ slots. If $(\pi_0,\pi_1)$ is the stationary distribution, then the steady-state-sampled expected persistence time of the GE chain is
\begin{equation}
\mathbb{E}[T_{\mathrm{GE}}]
= D\left(\pi_0\frac{1}{p_{01}} + \pi_1\frac{1}{p_{10}}\right).
\label{eq:etge-def}
\end{equation}
This definition is convenient because it is expressed directly in terms of the Markov parameters and has a clear operational meaning: sample the link in steady state, then ask how long the current binary link condition is expected to persist.

\paragraph{Remark: individual dwell times.}
For asymmetric thresholds ($S \ne 0$), the transition probabilities $p_{01}$ and $p_{10}$ differ, and the individual expected dwell times $D/p_{01}$ (mean time in state~0) and $D/p_{10}$ (mean time in state~1) may be of separate interest.  The weighted average $\mathbb{E}[T_{\mathrm{GE}}]$ in \eqref{eq:etge-def} is a single summary statistic; for applications where the asymmetry matters (e.g., mean outage duration vs.\ mean good-channel duration), the individual dwell times should be reported separately.

%% ====================================================================
\section{Closed-form GE parameters}
\label{sec:closedform}
%% ====================================================================
To derive the GE parameters, define
\[
X_0 \triangleq X(0),
\qquad
X_1 \triangleq X(D).
\]
Because $X(t)$ is stationary Gaussian with kernel \eqref{eq:kernel}, the pair $(X_0,X_1)$ is jointly Gaussian with zero mean and covariance matrix
\begin{equation}
\Sigma = \sigma^2
\begin{bmatrix}
1 & \rho\\
\rho & 1
\end{bmatrix},
\qquad
\rho \;\triangleq\; \frac{K(D)}{K(0)}.
\label{eq:sigma-rho}
\end{equation}
Thus $\rho$ is exactly the correlation coefficient between two adjacent slot-boundary samples.  For the two kernels considered here,
\begin{equation}
\rho_{\mathrm{sqexp}} = e^{-D^2/T_c^2},
\qquad
\rho_{\mathrm{exp}} = e^{-D/T_c}.
\label{eq:rho-both}
\end{equation}

Normalize the pair by defining
\[
Z_0 = \frac{X_0}{\sigma},
\qquad
Z_1 = \frac{X_1}{\sigma},
\qquad
s = \frac{S}{\sigma}.
\]
Then $(Z_0,Z_1)$ is standard bivariate normal with correlation $\rho$. Let $\Phi(\cdot)$ denote the standard normal CDF, let $\phi(\cdot)$ denote the standard normal PDF, and let $\Phi_2(a,b;\rho)$ denote the standard bivariate normal CDF with correlation $\rho$.

\subsection{General transition probabilities}

The first transition probability is
\begin{align}
p_{01}
&= \Pr(B(1)=1\mid B(0)=0) \nonumber\\
&= \Pr(X_1\ge S\mid X_0<S) \nonumber\\
&= \Pr(Z_1\ge s\mid Z_0<s) \nonumber\\
&= \frac{\Pr(Z_0<s,\, Z_1\ge s)}{\Pr(Z_0<s)}.
\label{eq:p01-start}
\end{align}
Now
\[
\Pr(Z_0<s)=\Phi(s),
\]
and
\begin{align*}
\Pr(Z_0<s,\, Z_1\ge s)
&= \Pr(Z_0<s) - \Pr(Z_0<s,\, Z_1<s)\\
&= \Phi(s) - \Phi_2(s,s;\rho).
\end{align*}
Hence
\begin{equation}
p_{01}
= \frac{\Phi(s)-\Phi_2(s,s;\rho)}{\Phi(s)}.
\label{eq:p01}
\end{equation}

Similarly,
\begin{align}
p_{10}
&= \Pr(B(1)=0\mid B(0)=1) \nonumber\\
&= \Pr(X_1<S\mid X_0\ge S) \nonumber\\
&= \Pr(Z_1<s\mid Z_0\ge s) \nonumber\\
&= \frac{\Pr(Z_0\ge s,\, Z_1<s)}{\Pr(Z_0\ge s)}.
\label{eq:p10-start}
\end{align}
The denominator is
\[
\Pr(Z_0\ge s)=1-\Phi(s),
\]
and the numerator is
\begin{align*}
\Pr(Z_0\ge s,\, Z_1<s)
&= \Pr(Z_1<s) - \Pr(Z_0<s,\, Z_1<s)\\
&= \Phi(s)-\Phi_2(s,s;\rho).
\end{align*}
Therefore
\begin{equation}
p_{10}
= \frac{\Phi(s)-\Phi_2(s,s;\rho)}{1-\Phi(s)}.
\label{eq:p10}
\end{equation}

It is useful to introduce the shorthand
\begin{equation}
q \triangleq \Phi(s),
\qquad
N \triangleq q - \Phi_2(s,s;\rho).
\label{eq:qN}
\end{equation}
Then \eqref{eq:p01}--\eqref{eq:p10} become simply
\begin{equation}
p_{01}=\frac{N}{q},
\qquad
p_{10}=\frac{N}{1-q}.
\label{eq:p-compact}
\end{equation}

We emphasize the distinction between what is exact and what is approximate.  The expressions \eqref{eq:p01}--\eqref{eq:p-compact} are \emph{exact} for the one-step conditional probabilities of the thresholded Gaussian process.  The matched first-order GE chain built from these parameters therefore preserves the exact one-step binary transitions.  However, it does not necessarily preserve the full higher-order memory of the binary process: run-length distributions, multi-step conditional probabilities, and other path-level statistics are approximated, not matched exactly.  Section~\ref{sec:markov-accuracy} quantifies this model mismatch.

\subsection{Closed-form GE expected persistence time as a function of $T_c$}
\label{sec:persistence-formula}
A two-state Markov chain with transition probabilities $p_{01},p_{10}$ has stationary distribution
\begin{equation}
\pi_0 = \frac{p_{10}}{p_{01}+p_{10}},
\qquad
\pi_1 = \frac{p_{01}}{p_{01}+p_{10}}.
\label{eq:pi-general}
\end{equation}
Substituting \eqref{eq:p-compact} into \eqref{eq:pi-general} gives
\begin{align*}
\pi_0
&= \frac{N/(1-q)}{N/q + N/(1-q)}
= \frac{1/(1-q)}{1/q + 1/(1-q)}
= q,
\\
\pi_1
&= 1-q.
\end{align*}
This is exactly what one would expect, since the stationary fraction of time spent in state $0$ is simply
\[
\Pr(X<S)=\Phi(s)=q,
\]
and the fraction spent in state $1$ is $1-q$.

Now substitute $\pi_0=q$, $\pi_1=1-q$, and \eqref{eq:p-compact} into \eqref{eq:etge-def}:
\begin{align}
\mathbb{E}[T_{\mathrm{GE}}]
&= D\left(q\frac{1}{N/q} + (1-q)\frac{1}{N/(1-q)}\right) \nonumber\\
&= D\left(\frac{q^2}{N} + \frac{(1-q)^2}{N}\right) \nonumber\\
&= D\,\frac{q^2+(1-q)^2}{N}.
\label{eq:etge-qn}
\end{align}
Replacing $q$ and $N$ by their definitions yields the closed form
\begin{equation}
\boxed{
\mathbb{E}[T_{\mathrm{GE}}]
= D\,
\frac{\Phi(s)^2 + \bigl(1-\Phi(s)\bigr)^2}
{\Phi(s)-\Phi_2(s,s;\rho)},
\qquad \rho = \frac{K(D)}{K(0)}.
}
\label{eq:exact-final}
\end{equation}
Equation \eqref{eq:exact-final} is the desired mapping from the one-step correlation~$\rho$ (and hence from~$T_c$ via~\eqref{eq:rho-both}) to the expected persistence time of the binary GE abstraction.  The formula applies to \emph{any} stationary Gaussian kernel; only the relationship between $\rho$ and $T_c$ is kernel-specific.

\subsection{Elementary closed form for the symmetric threshold}
\label{sec:arcsin}

When $S=0$ (i.e., $s=0$), the bivariate normal CDF $\Phi_2(0,0;\rho)$ admits an evaluation in terms of the arcsine function alone.  The following geometric argument yields the result directly.

Write the Cholesky factorization
\[
Z_0 = W_1,
\qquad
Z_1 = \rho\, W_1 + \sqrt{1-\rho^2}\,W_2,
\]
where $W_1,W_2\sim\mathcal{N}(0,1)$ are independent.  Their joint density is rotationally symmetric, so the probability of any event defined by half-planes through the origin equals the angular fraction of the plane that the event occupies.  In polar coordinates $(W_1,W_2) = (r\cos\theta,\, r\sin\theta)$, the event $\{Z_0>0,\;Z_1>0\}$ becomes the pair of inequalities
\[
\cos\theta > 0
\qquad\text{and}\qquad
\rho\cos\theta + \sqrt{1-\rho^2}\,\sin\theta > 0.
\]
Setting $\rho=\sin\psi$ with $\psi=\arcsin\rho$, the second condition reads $\sin(\psi+\theta)>0$, i.e., $\theta > -\arcsin\rho$.  Together with $\theta\in(-\pi/2,\,\pi/2)$ from the first condition, the feasible region is an arc of angular width $\pi/2+\arcsin\rho$.  Because probability equals angular fraction,
\begin{equation}
\Pr(Z_0>0,\;Z_1>0)
= \frac{\frac\pi2+\arcsin\rho}{2\pi}
= \frac{1}{4}+\frac{1}{2\pi}\arcsin\rho.
\label{eq:sheppard}
\end{equation}

\paragraph{Remark.}
Identity \eqref{eq:sheppard} is a classical result due to Sheppard~\cite{sheppard1900}, who established it in 1899 in the context of correlation measurement for dichotomized variables.  Our contribution is the application of this identity to derive elementary, closed-form GE transition probabilities and persistence times in the communications context.

The threshold-crossing numerator for $s=0$ follows immediately:
\[
N
= \Pr(Z_0<0,\;Z_1\ge0)
= \tfrac{1}{2} - \bigl(\tfrac{1}{4}+\tfrac{1}{2\pi}\arcsin\rho\bigr)
= \tfrac{1}{4} - \frac{1}{2\pi}\arcsin\rho.
\]
Since $q=\Phi(0)=\frac{1}{2}$ and $p_{01}=N/q$, $p_{10}=N/(1-q)$, both transition probabilities coincide:
\begin{equation}
\boxed{
p_{01} = p_{10}
= \frac{1}{2} - \frac{1}{\pi}\arcsin\rho,
\qquad S = 0.
}
\label{eq:arcsin-p}
\end{equation}
Substituting into \eqref{eq:etge-qn} with $q=\frac{1}{2}$ gives the expected GE persistence time in elementary form:
\begin{equation}
\boxed{
\mathbb{E}[T_{\mathrm{GE}}]
= \frac{2D}{\displaystyle 1-\frac{2}{\pi}\arcsin\rho},
\qquad S=0.
}
\label{eq:arcsin-etge}
\end{equation}
Equations~\eqref{eq:arcsin-p} and~\eqref{eq:arcsin-etge} involve only elementary functions and exactly reproduce the one-step transition probabilities for any kernel, with $\rho$ given by~\eqref{eq:rho-both}.  One can verify the limiting behavior directly: when $T_c\to0$, $\rho\to0$ and $p_{01}\to\frac{1}{2}$, so $\mathbb{E}[T_{\mathrm{GE}}]\to 2D$ (independent coin flips); when $T_c\to\infty$, $\rho\to1$ and $p_{01}\to0$, so $\mathbb{E}[T_{\mathrm{GE}}]\to\infty$ (frozen channel).

\subsection{Exact reduction via Owen's $T$-function for general threshold}
\label{sec:owen}

For general threshold $S \ne 0$, no elementary reduction analogous to the arcsine formula is available.  However, the equal-threshold bivariate normal probability admits an exact reduction in terms of Owen's $T$-function~\cite{owen1956}, which is substantially more explicit than leaving the answer in terms of~$\Phi_2$.

Define
\begin{equation}
a \;\triangleq\; \sqrt{\frac{1-\rho}{1+\rho}},
\label{eq:owen-a}
\end{equation}
and recall Owen's $T$-function
\begin{equation}
T(h,a)
\;\triangleq\;
\frac{1}{2\pi}
\int_0^a
\frac{\exp\!\bigl(-\tfrac{1}{2} h^2(1+t^2)\bigr)}{1+t^2}\,dt.
\label{eq:owen-def}
\end{equation}
For $0 \le \rho < 1$, the equal-threshold bivariate normal CDF satisfies the exact identity
\begin{equation}
\Phi_2(s,s;\rho) = \Phi(s) - 2\,T(s,a).
\label{eq:phi2-owen}
\end{equation}
Hence the threshold-crossing numerator reduces to $N = 2\,T(s,a)$, and the GE transition probabilities become
\begin{equation}
\boxed{
p_{01}
=
\frac{2\,T\!\left(s,\,\sqrt{\frac{1-\rho}{1+\rho}}\right)}{\Phi(s)},
\qquad
p_{10}
=
\frac{2\,T\!\left(s,\,\sqrt{\frac{1-\rho}{1+\rho}}\right)}{\Phi(-s)}.
}
\label{eq:owen-ge}
\end{equation}
The steady-state persistence time is therefore
\begin{equation}
\mathbb{E}[T_{\mathrm{GE}}]
=
D\,
\frac{\Phi(s)^2 + \bigl(1-\Phi(s)\bigr)^2}
{2\,T\!\left(s,\,\sqrt{\frac{1-\rho}{1+\rho}}\right)}.
\label{eq:owen-persistence}
\end{equation}

\paragraph{Recovery of the $S=0$ arcsine formula.}
The elementary result~\eqref{eq:arcsin-p} is the special case $s=0$.  Since $T(0,a)=\frac{1}{2\pi}\arctan(a)$,
\[
N = 2\,T(0,a)=\frac{1}{\pi}\arctan\!\sqrt{\frac{1-\rho}{1+\rho}}.
\]
The trigonometric identity $\arctan\!\sqrt{(1-\rho)/(1+\rho)} = \frac{\pi}{4}-\frac{1}{2}\arcsin(\rho)$ then gives $N = \frac{1}{4}-\frac{1}{2\pi}\arcsin(\rho)$, recovering $p_{01}=p_{10}=\frac{1}{2}-\frac{1}{\pi}\arcsin\rho$ exactly.  Thus the $S=0$ arcsine law is the elementary specialization of an exact all-threshold formula in one standard special function.

\paragraph{Remark.}
Owen's $T$-function is available in standard numerical libraries (e.g., \texttt{scipy.special.owens\_t} in Python) and admits rapidly convergent series and asymptotic expansions, making~\eqref{eq:owen-ge}--\eqref{eq:owen-persistence} convenient for both numerical evaluation and analytical approximation.

%% ====================================================================
\section{Accuracy of the first-order Markov approximation}
\label{sec:markov-accuracy}
%% ====================================================================

The matched first-order GE chain preserves the exact one-step binary transitions, but not necessarily the full higher-order memory.  This section quantifies the resulting model mismatch using two complementary diagnostics: the \emph{Markov gap} (a local, one-step conditional measure) and the \emph{run-length total-variation distance} (a global, path-shape measure).  We show that these diagnostics can trend in opposite directions, explain why mathematically, and identify practical regimes where the first-order approximation suffices.

\subsection{The Markov gap}

Define the \emph{Markov gap} as
\begin{equation}
\Delta_{\mathrm{gap}}(i,j) \;\triangleq\;
\bigl|\Pr(B_{n+1}=1 \mid B_n=j,\, B_{n-1}=i)
- \Pr(B_{n+1}=1 \mid B_n=j)\bigr|,
\label{eq:markov-gap}
\end{equation}
for each $(i,j) \in \{0,1\}^2$.  If the binary process were exactly first-order Markov, this gap would be zero for all $(i,j)$.

Figure~\ref{fig:markov-gap}(a) plots $\max_{i,j}\Delta_{\mathrm{gap}}(i,j)$ versus $T_c/D$ for both kernels with $S=0$.  The two kernels exhibit \emph{opposite} trends: the squared-exponential kernel's gap decreases monotonically, while the exponential kernel's gap increases.

\subsection{Why the two kernels exhibit opposite Markov-gap trends}
\label{sec:markov-gap-explanation}

\paragraph{Squared-exponential kernel: Gaussian non-Markovity shrinks.}
The deviation of the underlying Gaussian process from first-order Markov structure can be measured by $\rho_2 - \rho_1^2$, where $\rho_k = K(kD)/K(0)$.  For the squared-exponential kernel,
\[
\rho_2 - \rho_1^2 = e^{-4D^2/T_c^2} - e^{-2D^2/T_c^2} = -\frac{2D^2}{T_c^2} + O\!\left(\frac{D^4}{T_c^4}\right).
\]
Thus the Gaussian non-Markovity itself shrinks as $T_c^{-2}$, and with it the informational value of $B(n{-}1)$ beyond $B(n)$.  This gives a concrete mathematical reason for the decreasing Markov gap.

\paragraph{Exponential kernel: thresholding amplifies the near-threshold distinction.}
Here $\rho_k = \rho_1^k$, so the underlying Gaussian is exactly Markov and all binary non-Markovity is created by thresholding.  Writing the discrete-time OU update $X_{n+1} = \rho X_n + \eta_n$ with $\eta_n \sim \mathcal{N}(0,\sigma^2(1-\rho^2))$, the next-step transition probability given the latent value is
\[
\Pr(B_{n+1}=0 \mid X_n = x,\; B_n=1) = \Phi\!\left(\frac{s - \rho\, x}{\sigma\sqrt{1-\rho^2}}\right).
\]
As $T_c$ grows, $\rho \to 1$ and the innovation standard deviation $\sigma\sqrt{1-\rho^2} \approx \sigma\sqrt{2D/T_c}$ shrinks, so the transition probability becomes increasingly sensitive to the exact value of~$x$.  The sensitivity $|\partial/\partial x|$ grows as $\sqrt{T_c/(2D)}/\sigma$.  Now, conditioning on $(B_{n-1}=0,\,B_n=1)$ (``just entered state~1'') constrains $X_n$ to lie near the threshold, while conditioning on $(B_{n-1}=1,\,B_n=1)$ (``survived in state~1'') biases $X_n$ farther from the threshold.  As $T_c$ increases and the mapping from $x$ to transition probability steepens, this distinction becomes more consequential, driving the binary Markov gap upward even though the latent Gaussian process is exactly Markov.

\subsection{Markov gap versus run-length fit}
\label{sec:diagnostic-reconciliation}

\begin{figure}[H]
    \centering
    \includegraphics[width=0.98\linewidth]{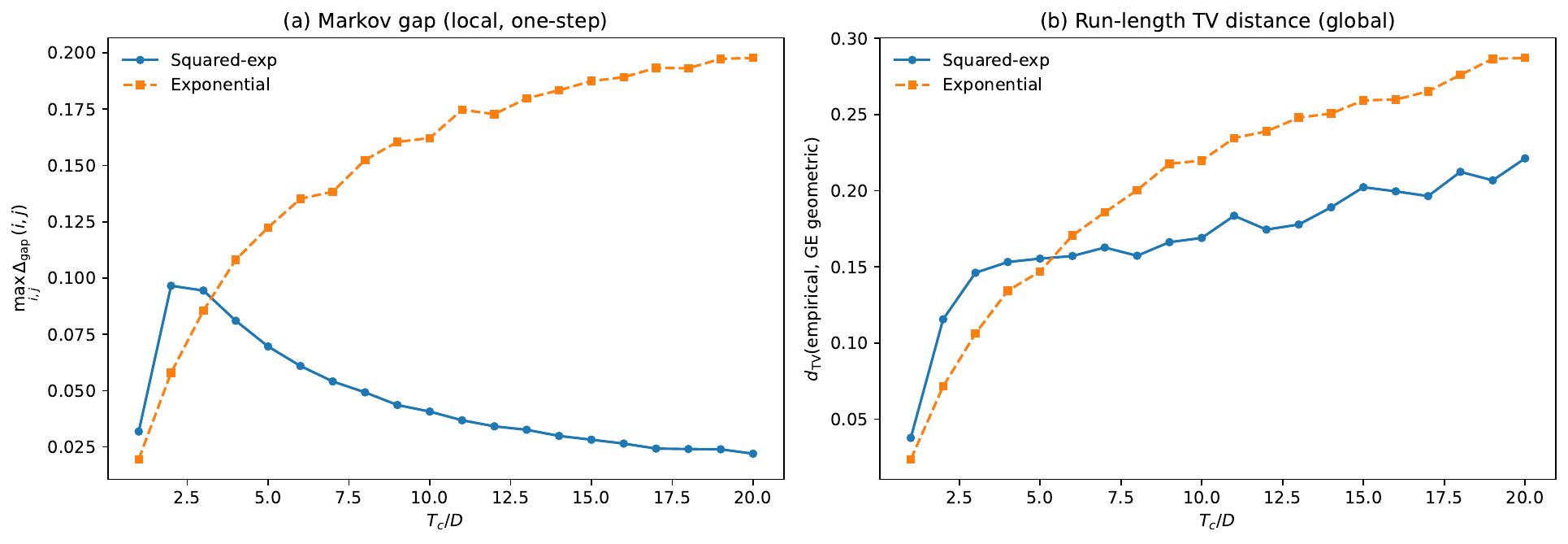}
    \caption{Two diagnostics for the matched first-order GE chain ($S/\sigma=0$, both kernels).  (a)~Maximum Markov gap: a local, one-step measure that \emph{decreases} for the squared-exponential kernel but \emph{increases} for the exponential kernel.  (b)~Run-length TV distance: a global, path-shape measure that increases with $T_c/D$ for both kernels.  The opposing trends in the squared-exponential case reflect the distinction between local conditional accuracy and global run-length fidelity.}
    \label{fig:markov-gap}
\end{figure}

Figure~\ref{fig:markov-gap}(b) shows a second diagnostic: the total-variation distance between the empirical run-length distribution and the GE geometric prediction.  Strikingly, for the squared-exponential kernel this TV distance \emph{increases} with $T_c/D$, the opposite of the Markov-gap trend in panel~(a).

This is not a contradiction.  The two diagnostics measure fundamentally different things:
\begin{itemize}[leftmargin=*]
\item The \textbf{Markov gap} is a \emph{local} conditional diagnostic.  It asks how much extra information $B(n{-}1)$ provides about the \emph{next symbol} beyond $B(n)$.
\item The \textbf{run-length TV distance} is a \emph{global} path-shape diagnostic.  It asks whether the \emph{entire run-length law} is close to geometric.
\end{itemize}
For the squared-exponential kernel, the one-step Markov gap shrinks because the Gaussian non-Markovity $|\rho_2 - \rho_1^2|$ vanishes as $T_c \to \infty$.  But the run-length TV distance grows because runs become longer (mean length $\sim T_c$), and within each run the exit probability varies systematically: it is elevated early in the run (the process is near the threshold) and depressed later (the process has drifted away from the threshold).  Even though the per-step deviation from the Markov prediction is small, this structured variation sculpts the run-length PMF away from geometric over the long duration of the run, producing heavier-than-geometric tails.

\paragraph{Shallow versus deep non-Markovity.}
A cross-kernel comparison sharpens this picture.  At large $T_c/D$, the exponential kernel has a \emph{larger} Markov gap than the squared-exponential kernel (Figure~\ref{fig:markov-gap}(a)), yet its run-length distribution is \emph{better} captured by a second-order Markov model (Figure~\ref{fig:ge-vs-bernoulli}).  The resolution is that non-Markovity can be \emph{shallow} or \emph{deep}.  For the exponential kernel, the latent Gaussian process is exactly first-order Markov; all binary non-Markovity is an artifact of thresholding a continuous Markov process into a binary alphabet, and it is concentrated at order~2.  One additional step of memory therefore absorbs most of the deviation.  For the squared-exponential kernel, the latent Gaussian has infinite-order correlations, so the induced binary non-Markovity is spread across many orders.  Each per-step deviation is small---hence the smaller Markov gap---but these deviations accumulate over long runs and resist correction by any low-order model.  In short, a large but shallow non-Markovity is easier to correct than a small but deep one.  Section~\ref{sec:run-lengths} confirms this numerically: the second-order model reduces $d_{\mathrm{TV}}$ from $0.196$ to $0.085$ for the exponential kernel at $T_c = 8$, but only from $0.175$ to $0.143$ for the squared-exponential kernel.

\paragraph{Practical guideline.}
For the squared-exponential kernel, the one-step GE approximation improves steadily with $T_c/D$, but the run-length fit degrades; a second-order Markov model is recommended when burst-length statistics matter (see Section~\ref{sec:run-lengths}).  For the exponential kernel, both diagnostics worsen with $T_c/D$, making the first-order GE chain most accurate at moderate values.  In all cases, the one-step persistence-time prediction $\mathbb{E}[T_{\mathrm{GE}}]$ remains accurate by construction.

%% ====================================================================
\section{Asymptotic scaling of the persistence time}
\label{sec:asymptotics}
%% ====================================================================
The empirical relation suggested by \eqref{eq:exact-final} is close to linear when $T_c$ is appreciably larger than the slot duration $D$. This can be explained analytically and, moreover, the asymptotic rate depends on the smoothness of the kernel at the origin.

\subsection{General asymptotic expansion}

From \eqref{eq:qN} and \eqref{eq:etge-qn}, the key quantity is the one-step threshold-crossing probability
\[
N = \Pr(Z_0<s,\, Z_1\ge s).
\]
Define the average and difference variables
\begin{equation}
M \triangleq \frac{Z_0+Z_1}{2},
\qquad
\Delta \triangleq Z_1-Z_0.
\label{eq:MDelta}
\end{equation}
Because $(Z_0,Z_1)$ is jointly Gaussian and symmetric, $(M,\Delta)$ is jointly Gaussian with
\begin{equation}
\mathrm{Var}(M)=\frac{1+\rho}{2},
\qquad
\mathrm{Var}(\Delta)=2(1-\rho),
\qquad
\mathrm{Cov}(M,\Delta)=0,
\label{eq:varMDelta}
\end{equation}
so $M$ and $\Delta$ are independent.

The crossing event $\{Z_0<s,\, Z_1\ge s\}$ is equivalent to
\[
\left\{M-\frac{\Delta}{2}<s,\; M+\frac{\Delta}{2}\ge s\right\}.
\]
If $\Delta<0$, this event cannot occur. If $\Delta\ge0$, it is equivalent to
\[
s-\frac{\Delta}{2} \le M < s+\frac{\Delta}{2}.
\]
Hence, conditioning on $\Delta$ and using independence,
\begin{equation}
N = \mathbb{E}\!\left[\mathbf{1}_{\{\Delta\ge0\}}\int_{s-\Delta/2}^{s+\Delta/2} f_M(m)\,dm\right],
\label{eq:N-exact-mdelta}
\end{equation}
where $f_M$ is the Gaussian density of $M$.

When $T_c\gg D$, one has $\rho\to 1$ and therefore $\Delta$ has small variance. The integration window in \eqref{eq:N-exact-mdelta} is then narrow, so to first order,
\[
\int_{s-\Delta/2}^{s+\Delta/2} f_M(m)\,dm
\approx f_M(s)\,\Delta.
\]
As $\rho\to1$, $f_M(s)=\phi(s)+O(1-\rho)$. Thus, to leading order,
\begin{equation}
N \approx \phi(s)\,\mathbb{E}[\Delta_+],
\label{eq:N-leading1}
\end{equation}
where $\Delta_+=\max\{\Delta,0\}$. Since $\Delta\sim \mathcal{N}(0,2(1-\rho))$,
\begin{equation}
\mathbb{E}[\Delta_+] = \frac{\sqrt{2(1-\rho)}}{\sqrt{2\pi}} = \sqrt{\frac{1-\rho}{\pi}}.
\label{eq:delta-plus}
\end{equation}
Combining \eqref{eq:N-leading1} and \eqref{eq:delta-plus} yields the key asymptotic result:

\begin{proposition}[Crossing probability asymptotics]
\label{prop:crossing}
Let $\varepsilon = 1 - \rho$.  Then
\begin{equation}
N = \phi(s)\sqrt{\frac{\varepsilon}{\pi}} + O(\varepsilon^{3/2}),
\label{eq:N-prop}
\end{equation}
and consequently
\begin{equation}
\mathbb{E}[T_{\mathrm{GE}}]
= \frac{\sqrt{\pi}\,\bigl(\Phi(s)^2 + (1-\Phi(s))^2\bigr)}{\phi(s)}\,\frac{D}{\sqrt{\varepsilon}} + O\!\left(\frac{D}{\varepsilon^{1/2}} \cdot \varepsilon\right).
\label{eq:etge-prop}
\end{equation}
\end{proposition}

The scaling law for $\mathbb{E}[T_{\mathrm{GE}}]$ in terms of $T_c$ depends on how $\varepsilon = 1-\rho$ relates to $T_c$, which is determined by the kernel.

\begin{corollary}[Squared-exponential kernel: linear scaling]
\label{cor:sqexp}
For $\rho = e^{-D^2/T_c^2}$, one has $\varepsilon = D^2/T_c^2 + O(D^4/T_c^4)$, and therefore
\begin{equation}
\mathbb{E}[T_{\mathrm{GE}}]
= \frac{\sqrt{\pi}\,\bigl(\Phi(s)^2 + (1-\Phi(s))^2\bigr)}{\phi(s)}\,T_c + O(D),
\qquad T_c \gg D.
\label{eq:linear-approx}
\end{equation}
\end{corollary}

\begin{corollary}[Exponential kernel: square-root scaling]
\label{cor:exp}
For $\rho = e^{-D/T_c}$, one has $\varepsilon = D/T_c + O(D^2/T_c^2)$, and therefore
\begin{equation}
\mathbb{E}[T_{\mathrm{GE}}]
= \frac{\sqrt{\pi}\,\bigl(\Phi(s)^2 + (1-\Phi(s))^2\bigr)}{\phi(s)}\,\sqrt{D\,T_c}
+ O\!\left(\sqrt{D^3/T_c}\right),
\qquad T_c \gg D.
\label{eq:sqrt-approx}
\end{equation}
\end{corollary}

The qualitative distinction is intuitive.  The squared-exponential kernel is infinitely differentiable at the origin, so its sample paths are smooth and the increment $X(D)-X(0)$ is $O(D/T_c)$; the resulting crossing probability scales as $D/T_c$, yielding linear growth of the persistence time.  The exponential kernel has a cusp at the origin (it is not differentiable there), so its increments are larger, of order $\sqrt{D/T_c}$, and the persistence time grows only as $\sqrt{T_c}$.  This dependence on kernel smoothness appears to be a new observation in the GE parameterization context.

\subsection{Direct recovery from the arcsin form}

For the symmetric case $s=0$ with the squared-exponential kernel, the same linear behavior follows in one line from \eqref{eq:arcsin-p}.  As $\rho\to1$, the identity $\arcsin\rho = \frac{\pi}{2} - \sqrt{2(1-\rho)} + O\bigl((1-\rho)^{3/2}\bigr)$ gives
\[
p_{01}
= \frac{1}{2} - \frac{1}{\pi}\!\left(\frac{\pi}{2} - \sqrt{2(1-\rho)}\,\right) + \cdots
= \frac{\sqrt{2(1-\rho)}}{\pi} + \cdots
\;\approx\; \frac{\sqrt{2}\,D}{\pi\,T_c},
\]
so $\mathbb{E}[T_{\mathrm{GE}}] = D/p_{01} \approx (\pi/\sqrt{2})\,T_c$, recovering \eqref{eq:linear-approx} for $s=0$ without the $M$-$\Delta$ machinery.

This approximation is also in the spirit of classical threshold-crossing arguments for Gaussian processes developed by Rice \cite{rice1944}.

%% ====================================================================
\section{Numerical results}
\label{sec:numerical}
%% ====================================================================
To instantiate the model concretely, we use $\sigma=1$, $S=0$, and $D=1$ throughout unless otherwise noted.  The choice $S=0$ makes the binary states symmetric around the mean in the baseline figures, while the full derivation above applies for arbitrary $S$.

\begin{figure}[H]
    \centering
    \includegraphics[width=0.95\linewidth]{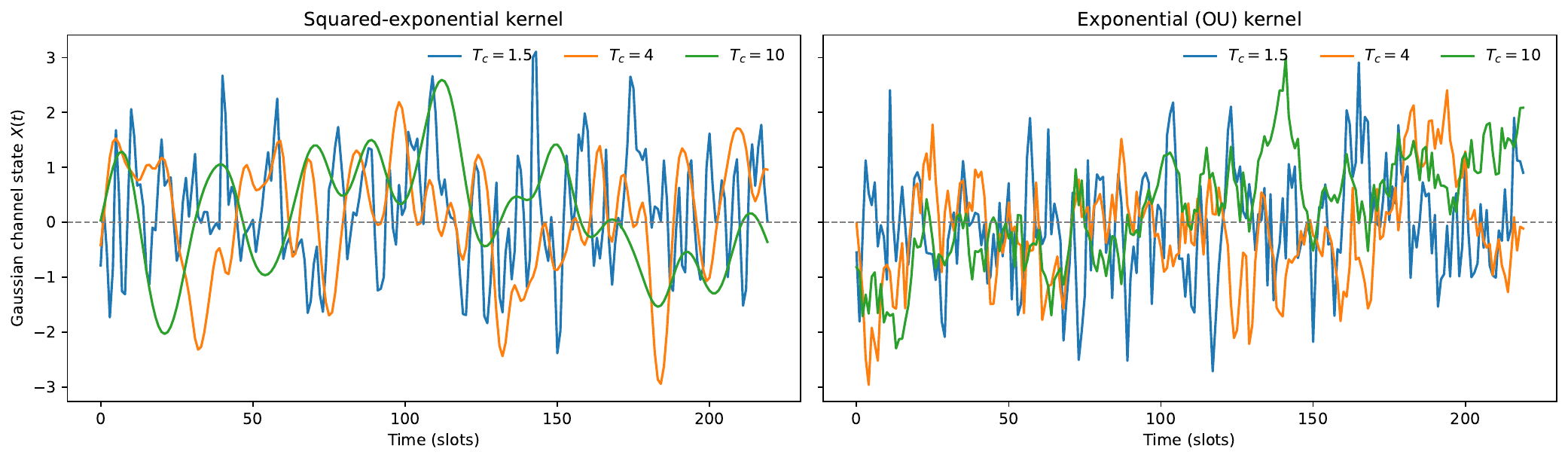}
    \caption{Sample paths of the stationary Gaussian fading process for three values of $T_c$.  Left: squared-exponential kernel~\eqref{eq:kernel} (smooth paths).  Right: exponential kernel~\eqref{eq:kernel-exp} (rougher paths due to the non-differentiable covariance at the origin).  Both panels use the same $T_c$ values and random seed to highlight the smoothness difference.}
    \label{fig:paths}
\end{figure}

\begin{figure}[H]
    \centering
    \includegraphics[width=0.92\linewidth]{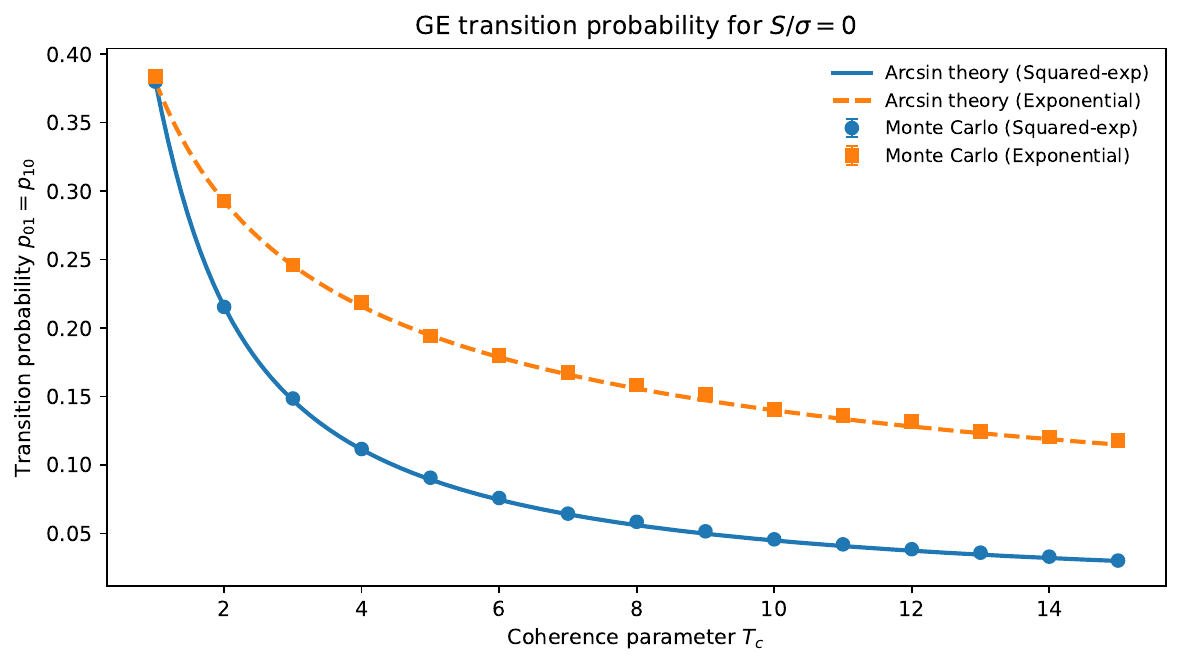}
    \caption{Matched GE transition probability $p_{01}=p_{10}$ versus $T_c$ for $S/\sigma=0$, both kernels. Solid and dashed curves are the arcsin formula~\eqref{eq:arcsin-p} with $\rho$ from~\eqref{eq:rho-both}; markers are Monte Carlo estimates with 95\% CIs. Note that the two kernels map the same $T_c$ to different values of $\rho$, so this figure compares GE parameters at matched $T_c$, not matched $\rho$.}
    \label{fig:trans-prob}
\end{figure}

\begin{figure}[H]
    \centering
    \includegraphics[width=0.92\linewidth]{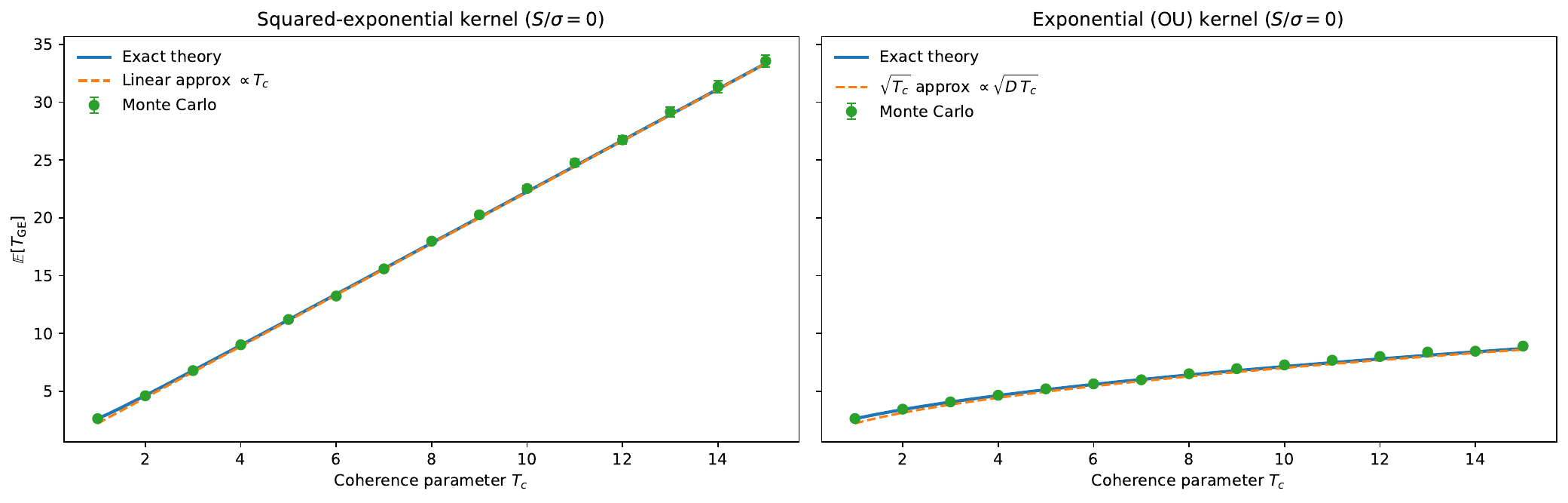}
    \caption{Expected persistence time $\mathbb{E}[T_{\mathrm{GE}}]$ versus $T_c$ for $S/\sigma=0$.  Left: squared-exponential kernel with exact theory~\eqref{eq:exact-final} (solid), linear asymptote~\eqref{eq:linear-approx} (dashed), and Monte Carlo (circles).  Right: exponential kernel with exact theory (solid), $\sqrt{T_c}$ asymptote~\eqref{eq:sqrt-approx} (dashed), and Monte Carlo (circles).  The contrasting growth rates (linear vs.\ $\sqrt{T_c}$) are clearly visible.}
    \label{fig:main}
\end{figure}

\begin{figure}[H]
    \centering
    \includegraphics[width=0.95\linewidth]{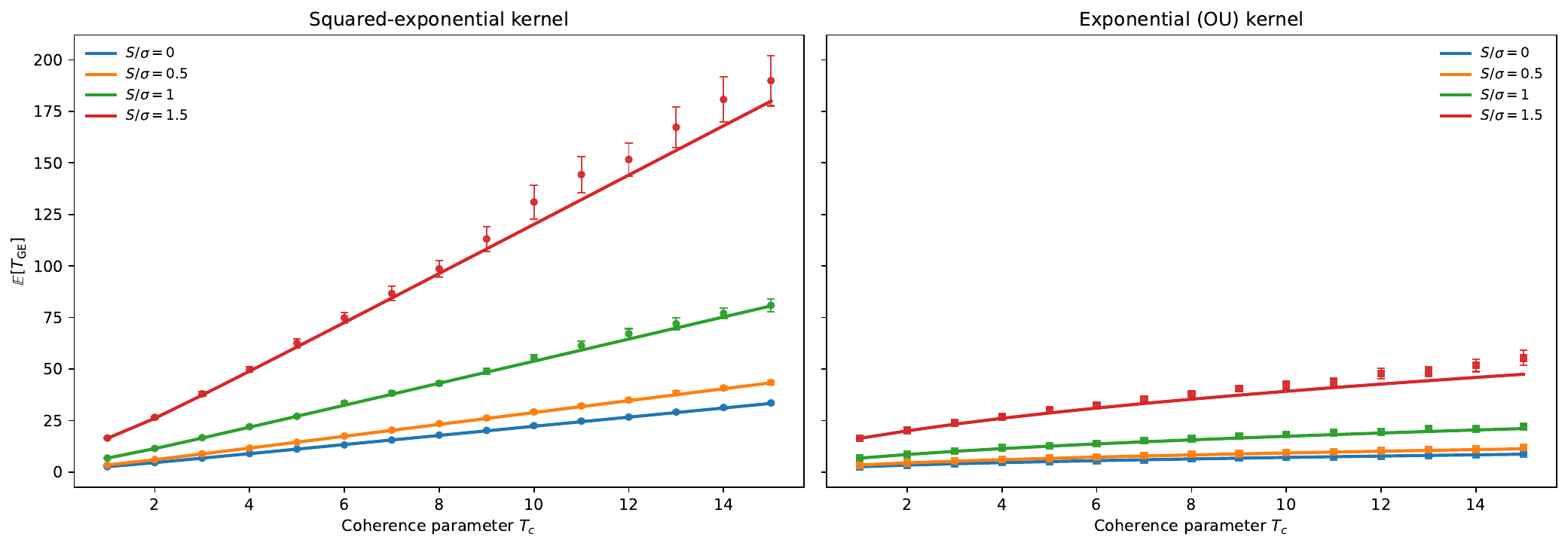}
    \caption{Effect of the threshold level on $\mathbb{E}[T_{\mathrm{GE}}]$, both kernels.  Left: squared-exponential.  Right: exponential.  Solid lines are the closed-form theory~\eqref{eq:exact-final}; markers show Monte Carlo means with 95\% CIs.  Only nonnegative thresholds are shown (the mapping is symmetric under $S/\sigma \mapsto -S/\sigma$).  The square-root scaling of the exponential kernel is evident in the sublinear curvature of the right panel.}
    \label{fig:multithreshold}
\end{figure}

All figures present both kernels side by side.  Figures~\ref{fig:paths}--\ref{fig:main} validate the closed-form expressions and contrast the linear versus $\sqrt{T_c}$ scaling.  Figure~\ref{fig:multithreshold} shows threshold effects for both kernels.  Figures~\ref{fig:dual-kernel}--\ref{fig:ge-vs-bernoulli} and Table~\ref{tab:summary} extend to asymmetric thresholds, run-length diagnostics, and a quantitative summary.

For the Monte Carlo estimates, we generated $250$ independent Gaussian sequences of length $1200$ for each $T_c$ on a grid, thresholded them via~\eqref{eq:binary-def}, and estimated transition probabilities from observed counts with a Jeffreys-style $1/2$ pseudo-count. Error bars denote 95\% confidence intervals over replications.

Figure~\ref{fig:trans-prob} validates the arcsin formula~\eqref{eq:arcsin-p} for both kernels: $p_{01}$ decreases monotonically from $\frac{1}{2}$ toward $0$ as $T_c$ increases, and the Monte Carlo estimates fall within tight confidence intervals of the theoretical curves.  At every $T_c > D$, the squared-exponential kernel yields a smaller $p_{01}$ (higher $\rho$) than the exponential kernel, consistent with the ordering $e^{-D^2/T_c^2} > e^{-D/T_c}$.

Figure~\ref{fig:main} contrasts the persistence-time scaling for the two kernels.  For the squared-exponential kernel (left panel), $\mathbb{E}[T_{\mathrm{GE}}]$ is nearly linear in $T_c$; for $S=0$ the slope is
\[
\mathbb{E}[T_{\mathrm{GE}}] \approx \frac{\pi}{\sqrt{2}}\,T_c \approx 2.22\,T_c,
\qquad T_c\gg D.
\]
For the exponential kernel (right panel), the growth follows $\sqrt{T_c}$ as predicted by Corollary~\ref{cor:exp}.  The visual contrast between linear and square-root growth is striking and directly reflects the kernel smoothness distinction identified in Section~\ref{sec:asymptotics}.

Figure~\ref{fig:multithreshold} shows that increasing the threshold magnitude increases the persistence time for all kernels, since a more extreme threshold produces a rarer crossing event.

\subsection{Dual-kernel comparison: transition probabilities and dwell times}
\label{sec:dual-kernel}

\begin{figure}[H]
    \centering
    \includegraphics[width=0.98\linewidth]{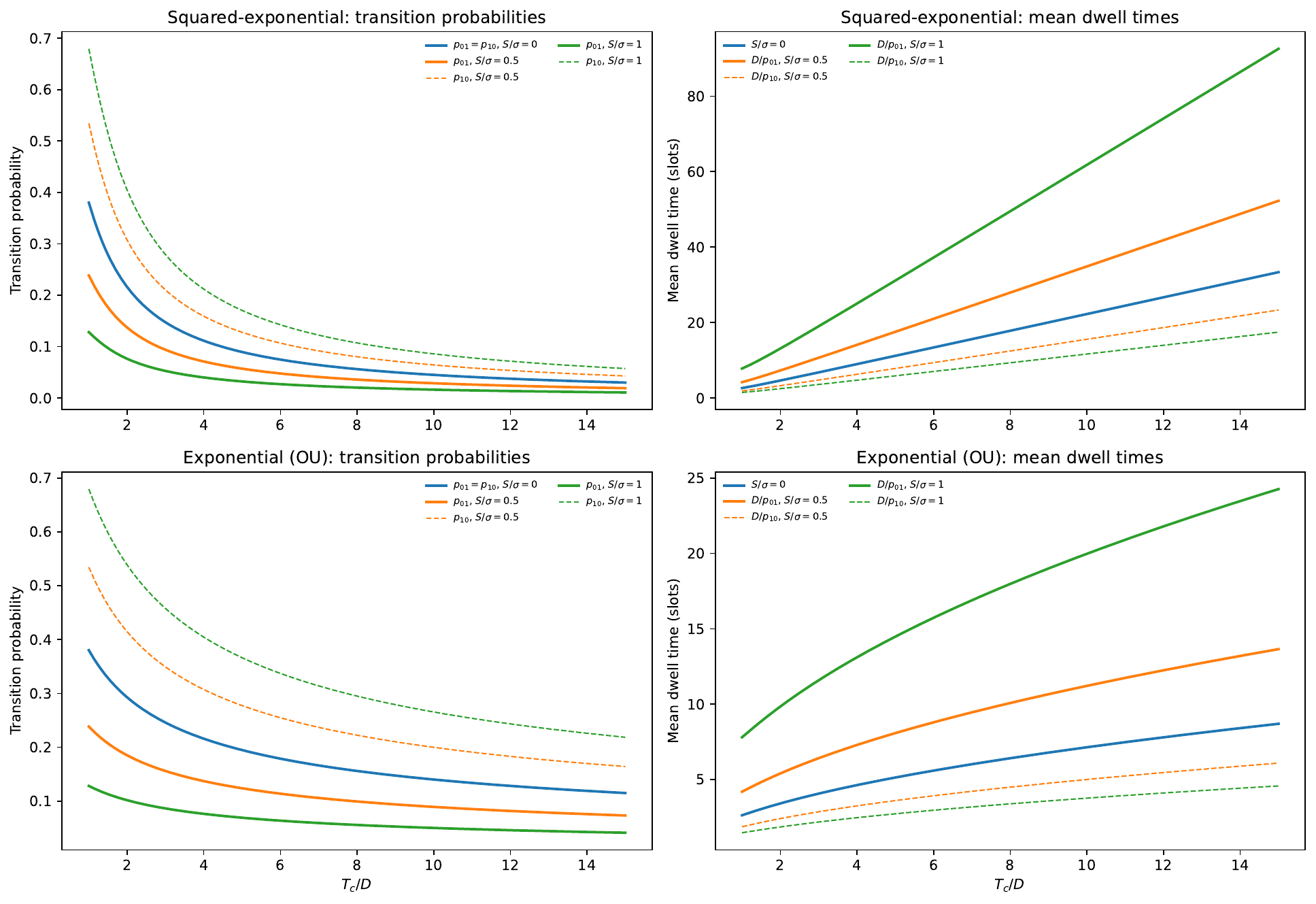}
    \caption{Matched GE transition probabilities (left column) and mean dwell times (right column) for three threshold levels $S/\sigma \in \{0, 0.5, 1.0\}$.  Top row: squared-exponential kernel.  Bottom row: exponential kernel.  For $S \ne 0$, $p_{01}$ (solid) and $p_{10}$ (dashed) differ.}
    \label{fig:dual-kernel}
\end{figure}

Figure~\ref{fig:dual-kernel} compares the matched GE transition probabilities and mean dwell times for both kernels across three threshold levels $S/\sigma \in \{0, 0.5, 1.0\}$.  Each row shows one kernel; each column shows transition probabilities (left) or mean dwell times (right).  For $S=0$ the two transition probabilities coincide; for $S>0$ the asymmetry is visible, with $p_{10} > p_{01}$.  Within each panel, the curves for both kernels are directly comparable: at the same $T_c/D > 1$, the squared-exponential kernel yields smaller transition probabilities and longer dwell times, reflecting its higher one-step correlation.

\subsection{Run-length distributions: GE, second-order Markov, and Bernoulli}
\label{sec:run-lengths}

\begin{figure}[H]
    \centering
    \includegraphics[width=0.95\linewidth]{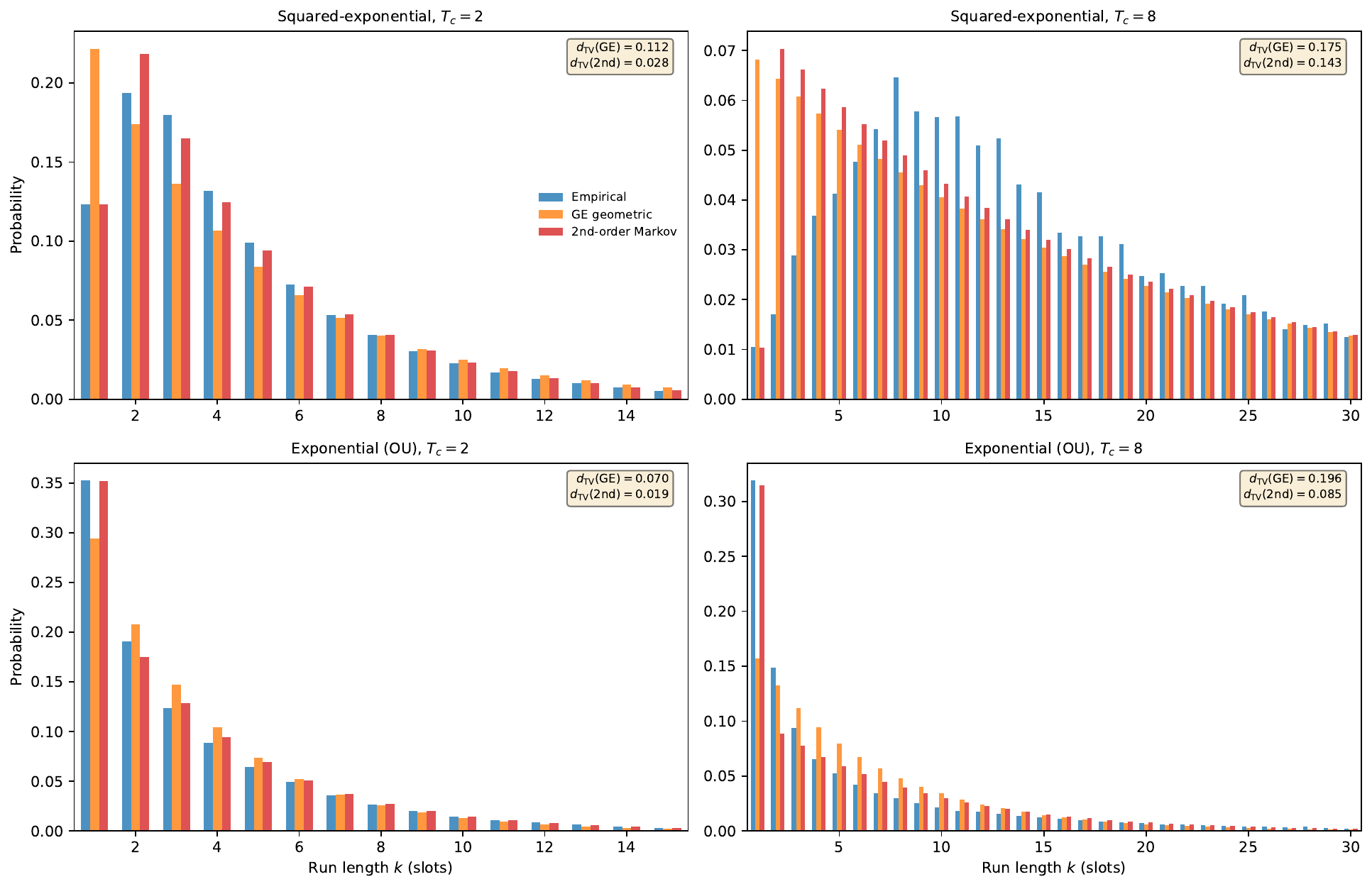}
    \caption{Run-length distributions for state-1 runs ($S/\sigma = 0$).  Top row: squared-exponential kernel.  Bottom row: exponential kernel.  Left column: $T_c = 2$.  Right column: $T_c = 8$.  Each panel annotates the total-variation distances $d_{\mathrm{TV}}$ for the first-order GE geometric and second-order Markov fits.  At $T_c = 2$ the exponential kernel already shows a better GE fit than the squared-exponential ($0.070$ vs.\ $0.112$).  At $T_c = 8$ both kernels develop heavier-than-geometric tails, but the second-order Markov correction is far more effective for the exponential kernel ($d_{\mathrm{TV}}$ drops from $0.196$ to $0.085$) than for the squared-exponential ($0.175$ to $0.143$), reflecting the shallower binary non-Markovity of the exponential kernel.}
    \label{fig:ge-vs-bernoulli}
\end{figure}

Figure~\ref{fig:ge-vs-bernoulli} compares empirical run-length distributions with the matched GE geometric prediction and a second-order Markov model, for both kernels at $T_c \in \{2, 8\}$.  At $T_c = 2$ (left column), the exponential kernel already produces a close GE fit ($d_{\mathrm{TV}} = 0.070$), while the squared-exponential kernel shows a larger deviation ($d_{\mathrm{TV}} = 0.112$).  At $T_c = 8$ (right column), both kernels develop heavier-than-geometric tails, with comparable first-order TV distances ($0.175$ for squared-exponential, $0.196$ for exponential).

The most striking cross-kernel difference is in how well the second-order Markov model corrects the fit.  For the exponential kernel, the second-order TV distance drops from $0.196$ to $0.085$ at $T_c = 8$, a substantial improvement.  For the squared-exponential kernel, the reduction is much smaller: from $0.175$ to $0.143$.  This contrast is consistent with the nature of the two kernels.  The exponential kernel's underlying Gaussian process is Markov, so the binary non-Markovity introduced by thresholding is relatively ``shallow'' and well captured by one additional step of memory.  The squared-exponential kernel has infinite-order Gaussian correlations, so the binary process retains higher-order structure that even two steps of memory cannot fully absorb.

\subsection{Quantitative summary}
\label{sec:summary-table}

Table~\ref{tab:summary} provides a systematic comparison across both kernels, three threshold levels, and five values of $T_c/D$.  For each configuration, we report the maximum Markov gap $\max_{i,j}\Delta_{\mathrm{gap}}(i,j)$, the total variation distance between the empirical and GE-predicted run-length distributions ($d_{\mathrm{TV}}$(GE)), the corresponding TV distance for the second-order Markov model ($d_{\mathrm{TV}}$(2nd)), and the relative error in $\mathbb{E}[T_{\mathrm{GE}}]$ between the closed-form prediction and the Monte Carlo estimate.

\begin{table}[H]
\centering
\caption{Quantitative summary of GE model accuracy across kernels, thresholds, and $T_c/D$.}
\label{tab:summary}
\small
\begin{tabular}{rrlrrrr}
\toprule
$T_c/D$ & $S/\sigma$ & Kernel & Max gap & $d_{\mathrm{TV}}$(GE) & $d_{\mathrm{TV}}$(2nd) & $\mathbb{E}[T_{\mathrm{GE}}]$ err.\% \\
\midrule
2 & 0 & SqExp & 0.0959 & 0.1156 & 0.0268 & 0.19 \\
2 & 0 & Exp & 0.0568 & 0.0715 & 0.0171 & 0.15 \\
2 & 0.5 & SqExp & 0.1136 & 0.1388 & 0.0219 & 0.71 \\
2 & 0.5 & Exp & 0.0672 & 0.0524 & 0.0100 & 0.12 \\
2 & 1 & SqExp & 0.1203 & 0.1516 & 0.0172 & 0.33 \\
2 & 1 & Exp & 0.0657 & 0.0408 & 0.0062 & 0.16 \\
5 & 0 & SqExp & 0.0697 & 0.1555 & 0.1090 & 0.20 \\
5 & 0 & Exp & 0.1232 & 0.1535 & 0.0627 & 0.74 \\
5 & 0.5 & SqExp & 0.0922 & 0.1915 & 0.1308 & 0.83 \\
5 & 0.5 & Exp & 0.1316 & 0.1253 & 0.0413 & 0.36 \\
5 & 1 & SqExp & 0.1177 & 0.2093 & 0.1399 & 0.53 \\
5 & 1 & Exp & 0.1273 & 0.1106 & 0.0320 & 0.78 \\
8 & 0 & SqExp & 0.0503 & 0.1674 & 0.1372 & 0.80 \\
8 & 0 & Exp & 0.1545 & 0.2038 & 0.0882 & 0.17 \\
8 & 0.5 & SqExp & 0.0656 & 0.1910 & 0.1540 & 0.85 \\
8 & 0.5 & Exp & 0.1564 & 0.1664 & 0.0676 & 2.01 \\
8 & 1 & SqExp & 0.0849 & 0.2246 & 0.1757 & 1.83 \\
8 & 1 & Exp & 0.1563 & 0.1473 & 0.0517 & 1.79 \\
10 & 0 & SqExp & 0.0396 & 0.1635 & 0.1378 & 0.14 \\
10 & 0 & Exp & 0.1661 & 0.2269 & 0.1024 & 0.20 \\
10 & 0.5 & SqExp & 0.0559 & 0.2026 & 0.1716 & 2.18 \\
10 & 0.5 & Exp & 0.1721 & 0.1881 & 0.0848 & 0.07 \\
10 & 1 & SqExp & 0.0738 & 0.2307 & 0.1909 & 0.09 \\
10 & 1 & Exp & 0.1717 & 0.1669 & 0.0621 & 1.46 \\
15 & 0 & SqExp & 0.0288 & 0.1831 & 0.1639 & 0.89 \\
15 & 0 & Exp & 0.1850 & 0.2551 & 0.1300 & 0.91 \\
15 & 0.5 & SqExp & 0.0390 & 0.2076 & 0.1873 & 0.78 \\
15 & 0.5 & Exp & 0.1901 & 0.2284 & 0.1058 & 0.37 \\
15 & 1 & SqExp & 0.0519 & 0.2271 & 0.2007 & 0.42 \\
15 & 1 & Exp & 0.1900 & 0.1987 & 0.0808 & 1.16 \\
\bottomrule
\end{tabular}

\end{table}

Key observations from Table~\ref{tab:summary}:
\begin{itemize}[leftmargin=*]
\item At small $T_c/D$ (e.g.\ $T_c/D = 2$), the exponential kernel shows smaller Markov gaps and TV distances than the squared-exponential kernel.  At larger $T_c/D$, the pattern reverses: the squared-exponential kernel's Markov gap decreases toward zero while the exponential kernel's gap increases, reflecting the different higher-order correlation structures.
\item The relative error in $\mathbb{E}[T_{\mathrm{GE}}]$ is uniformly small (typically below $2\%$) for both kernels, since the persistence-time formula depends only on one-step statistics that the GE chain matches by construction.
\item The second-order Markov model ($d_{\mathrm{TV}}$(2nd)) consistently improves over the first-order GE model ($d_{\mathrm{TV}}$(GE)) for both kernels, confirming that two steps of memory capture meaningful additional structure.
\end{itemize}

%% ====================================================================
\section{Conclusion}
%% ====================================================================
The correlated Gaussian process and the Gilbert-Elliott chain are the standard channel models at the physical and link layers, respectively.  This paper has provided the exact, closed-form bridge between them: given any stationary Gaussian fading model, the GE transition probabilities and persistence times follow in closed form from the one-step correlation coefficient alone.

The main results:
\begin{itemize}[leftmargin=*]
\item \textbf{Exact cross-layer bridge.}  The GE transition probabilities are given in closed form by~\eqref{eq:exact-final}, reducing to Owen's $T$-function for general thresholds~\eqref{eq:owen-ge} and to the elementary arcsine identity for $S=0$~\eqref{eq:arcsin-p}.
\item \textbf{Kernel-dependent scaling.}  The persistence time $\mathbb{E}[T_{\mathrm{GE}}]$ grows linearly in~$T_c$ for the squared-exponential kernel but as~$\sqrt{T_c}$ for the exponential kernel, governed by kernel smoothness at the origin (Proposition~\ref{prop:crossing} and Corollaries~\ref{cor:sqexp}--\ref{cor:exp}).
\item \textbf{Diagnostic reconciliation.}  Two model-mismatch diagnostics, the one-step Markov gap and the run-length TV distance, can trend in opposite directions, reflecting the distinction between local conditional accuracy and global path-shape fidelity.  For the exponential kernel, thresholding amplifies the near-threshold distinction as $T_c$ grows, driving the Markov gap upward despite exact Markov structure in the latent process.
\item \textbf{Practical regime.}  The matched first-order GE chain is useful across all regimes for one-step transition modeling and persistence-time prediction, but is not a faithful generator of long-run statistics when $T_c/D$ is large.  A second-order Markov model captures the heavier-than-geometric tails that emerge in that regime.
\end{itemize}

\section*{Acknowledgment}
The development of the mathematical results, code for plots, and writing of this paper have involved the use of AI tools/agents including Claude Code and ChatGPT.  The human authors accept full responsibility for its contents.

\bibliographystyle{ieeetr}
\bibliography{refs}

@article{owen1956,
  author  = {Owen, Donald B.},
  title   = {Tables for Computing Bivariate Normal Probabilities},
  journal = {The Annals of Mathematical Statistics},
  volume  = {27},
  number  = {4},
  pages   = {1075--1090},
  year    = {1956},
  month   = dec,
  doi     = {10.1214/aoms/1177728074},
  url     = {https://doi.org/10.1214/aoms/1177728074}
}

@article{gilbert1960,
  author  = {Gilbert, Edgar N.},
  title   = {Capacity of a Burst-Noise Channel},
  journal = {The Bell System Technical Journal},
  volume  = {39},
  number  = {5},
  pages   = {1253--1265},
  year    = {1960},
  month   = sep,
  doi     = {10.1002/j.1538-7305.1960.tb03959.x},
  url     = {https://doi.org/10.1002/j.1538-7305.1960.tb03959.x}
}

@article{elliott1963,
  author  = {Elliott, E. O.},
  title   = {Estimates of Error Rates for Codes on Burst-Noise Channels},
  journal = {The Bell System Technical Journal},
  volume  = {42},
  number  = {5},
  pages   = {1977--1997},
  year    = {1963},
  month   = sep,
  doi     = {10.1002/j.1538-7305.1963.tb00955.x},
  url     = {https://doi.org/10.1002/j.1538-7305.1963.tb00955.x}
}

@article{rice1944,
  author  = {Rice, S. O.},
  title   = {Mathematical Analysis of Random Noise},
  journal = {The Bell System Technical Journal},
  volume  = {23},
  number  = {3},
  pages   = {282--332},
  year    = {1944},
  month   = jul,
  doi     = {10.1002/j.1538-7305.1944.tb00874.x},
  url     = {https://doi.org/10.1002/j.1538-7305.1944.tb00874.x}
}

@article{clarke1968,
  author  = {Clarke, R. H.},
  title   = {A Statistical Theory of Mobile-Radio Reception},
  journal = {The Bell System Technical Journal},
  volume  = {47},
  number  = {6},
  pages   = {957--1000},
  year    = {1968},
  month   = jul,
  doi     = {10.1002/j.1538-7305.1968.tb00069.x},
  url     = {https://doi.org/10.1002/j.1538-7305.1968.tb00069.x}
}

@book{jakes1974,
  editor    = {Jakes, William C.},
  title     = {Microwave Mobile Communications},
  publisher = {John Wiley \& Sons},
  address   = {New York, NY, USA},
  year      = {1974},
  note      = {Often cited as the Clarke/Jakes fading reference; includes classical Doppler spectrum and correlation models}
}

@article{kanal_sastry1978,
  author  = {Kanal, Laveen N. and Sastry, Ark},
  title   = {Models for Channels with Memory and Their Applications to Error Control},
  journal = {Proceedings of the IEEE},
  volume  = {66},
  number  = {7},
  pages   = {724--744},
  year    = {1978},
  month   = jul,
  doi     = {10.1109/PROC.1978.11013},
  url     = {https://doi.org/10.1109/PROC.1978.11013}
}

@article{sadeghi2008fsmc_survey,
  author  = {Sadeghi, Parastoo and Kennedy, Rodney A. and Rapajic, Predrag B. and Shams, Ramtin},
  title   = {Finite-State Markov Modeling of Fading Channels: {A} Survey of Principles and Applications},
  journal = {IEEE Signal Processing Magazine},
  volume  = {25},
  number  = {5},
  pages   = {57--80},
  year    = {2008},
  month   = sep,
  doi     = {10.1109/MSP.2008.926683},
  url     = {https://doi.org/10.1109/MSP.2008.926683}
}

@article{wangmoayeri1995fsmc,
  author  = {Wang, Hong Shen and Moayeri, Nader},
  title   = {Finite-State Markov Channel---A Useful Model for Radio Communication Channels},
  journal = {IEEE Transactions on Vehicular Technology},
  volume  = {44},
  number  = {1},
  pages   = {163--171},
  year    = {1995},
  month   = feb,
  doi     = {10.1109/25.350282},
  url     = {https://doi.org/10.1109/25.350282}
}

@article{wangchang1996verify,
  author  = {Wang, Hong Shen and Chang, Pao-Chi},
  title   = {On Verifying the First-Order Markovian Assumption for a Rayleigh Fading Channel Model},
  journal = {IEEE Transactions on Vehicular Technology},
  volume  = {45},
  number  = {2},
  pages   = {353--357},
  year    = {1996},
  month   = may,
  doi     = {10.1109/25.492909},
  url     = {https://doi.org/10.1109/25.492909}
}

@article{zhangkassam1999fsmc,
  author  = {Zhang, Q. and Kassam, S. A.},
  title   = {Finite-State Markov Model for Rayleigh Fading Channels},
  journal = {IEEE Transactions on Communications},
  volume  = {47},
  number  = {11},
  pages   = {1688--1692},
  year    = {1999},
  month   = nov,
  doi     = {10.1109/26.803503},
  url     = {https://doi.org/10.1109/26.803503}
}

@article{tanbeaulieu2000firstorder,
  author  = {Tan, Christopher C. and Beaulieu, Norman C.},
  title   = {On First-Order Markov Modeling for the Rayleigh Fading Channel},
  journal = {IEEE Transactions on Communications},
  volume  = {48},
  number  = {12},
  pages   = {2032--2040},
  year    = {2000},
  month   = dec,
  doi     = {10.1109/26.891214},
  url     = {https://doi.org/10.1109/26.891214}
}

@article{zorzi1997arq,
  author  = {Zorzi, Michele and Rao, R. R. and Milstein, L. B.},
  title   = {ARQ Error Control for Fading Mobile Radio Channels},
  journal = {IEEE Transactions on Vehicular Technology},
  volume  = {46},
  number  = {2},
  pages   = {445--455},
  year    = {1997},
  month   = may,
  doi     = {10.1109/25.580783},
  url     = {https://doi.org/10.1109/25.580783}
}

@inproceedings{zorzi1995accuracy,
  author    = {Zorzi, Michele and Rao, R. R. and Milstein, L. B.},
  title     = {On the Accuracy of a First-Order Markov Model for Data Transmission on Fading Channels},
  booktitle = {Proceedings of ICUPC '95 -- 4th IEEE International Conference on Universal Personal Communications},
  address   = {Tokyo, Japan},
  month     = nov,
  year      = {1995},
  pages     = {211--215},
  doi       = {10.1109/ICUPC.1995.496890},
  url       = {https://doi.org/10.1109/ICUPC.1995.496890}
}

@inproceedings{zunigakrishnamachari2004transitional,
  author    = {Zuniga, Marco and Krishnamachari, Bhaskar},
  title     = {Analyzing the Transitional Region in Low Power Wireless Links},
  booktitle = {Proceedings of the First Annual IEEE Communications Society Conference on Sensor and Ad Hoc Communications and Networks (SECON 2004)},
  address   = {Santa Clara, CA, USA},
  year      = {2004},
  pages     = {517--526},
  doi       = {10.1109/SAHCN.2004.1381954},
  url       = {https://doi.org/10.1109/SAHCN.2004.1381954}
}

@inproceedings{srinivasan2008betafactor,
  author    = {Srinivasan, Kannan and Kazandjieva, Maria A. and Agarwal, Saatvik and Levis, Philip},
  title     = {The {$\beta$}-Factor: Measuring Wireless Link Burstiness},
  booktitle = {Proceedings of the 6th ACM Conference on Embedded Network Sensor Systems (SenSys '08)},
  address   = {Raleigh, NC, USA},
  year      = {2008},
  pages     = {29--42},
  doi       = {10.1145/1460412.1460416},
  url       = {https://doi.org/10.1145/1460412.1460416}
}

@article{liuzhougiannakis2005queue_amc,
  author  = {Liu, Q. and Zhou, S. and Giannakis, G. B.},
  title   = {Queuing with Adaptive Modulation and Coding over Wireless Links: Cross-Layer Analysis and Design},
  journal = {IEEE Transactions on Wireless Communications},
  volume  = {4},
  number  = {3},
  pages   = {1142--1152},
  year    = {2005},
  month   = may,
  doi     = {10.1109/TWC.2005.847005},
  url     = {https://doi.org/10.1109/TWC.2005.847005}
}

@article{moltchanov2006crosslayer,
  author  = {Moltchanov, Dmitri and Koucheryavy, Yevgeni and Harju, Jarmo},
  title   = {Cross-layer Modeling of Wireless Channels for Data-link and IP Layer Performance Evaluation},
  journal = {Computer Communications},
  volume  = {29},
  number  = {7},
  pages   = {827--841},
  year    = {2006},
  month   = apr,
  doi     = {10.1016/j.comcom.2005.08.005},
  url     = {https://doi.org/10.1016/j.comcom.2005.08.005}
}

@inproceedings{kimkrunz1999arq_delay,
  author    = {Kim, Jeong Geun and Krunz, Marwan},
  title     = {Delay Analysis of Selective Repeat ARQ for a Markovian Source over a Wireless Channel},
  booktitle = {Proceedings of the 2nd ACM International Workshop on Wireless Mobile Multimedia (WoWMoM '99)},
  address   = {Seattle, WA, USA},
  year      = {1999},
  pages     = {59--66},
  doi       = {10.1145/313256.313278},
  url       = {https://doi.org/10.1145/313256.313278}
}

@article{sinopoli2004intermittent,
  author  = {Sinopoli, Bruno and Schenato, Luca and Franceschetti, Massimo and Poolla, Kameshwar and Jordan, Michael I. and Sastry, Shankar S.},
  title   = {Kalman Filtering with Intermittent Observations},
  journal = {IEEE Transactions on Automatic Control},
  volume  = {49},
  number  = {9},
  pages   = {1453--1464},
  year    = {2004},
  month   = sep,
  doi     = {10.1109/TAC.2004.834121},
  url     = {https://doi.org/10.1109/TAC.2004.834121}
}

@article{huangdey2007markov_losses,
  author  = {Huang, Minyi and Dey, Subhrakanti},
  title   = {Stability of Kalman Filtering with Markovian Packet Losses},
  journal = {Automatica},
  volume  = {43},
  number  = {4},
  pages   = {598--607},
  year    = {2007},
  month   = apr,
  doi     = {10.1016/j.automatica.2006.10.023},
  url     = {https://doi.org/10.1016/j.automatica.2006.10.023}
}

@article{wushiandersonjohansson2017ge_kalman,
  author  = {Wu, Junfeng and Shi, Guodong and Anderson, Brian D. O. and Johansson, Karl Henrik},
  title   = {Kalman Filtering over Gilbert--Elliott Channels: Stability Conditions and the Critical Curve},
  journal = {IEEE Transactions on Automatic Control},
  year    = {2017},
  doi     = {10.1109/TAC.2017.2732821},
  url     = {https://doi.org/10.1109/TAC.2017.2732821},
  note    = {Published version of earlier arXiv:1411.1217}
}

@article{chakravortymahajan2020remote_est,
  author  = {Chakravorty, Jhelum and Mahajan, Aditya},
  title   = {Remote Estimation Over a Packet-Drop Channel With Markovian State},
  journal = {IEEE Transactions on Automatic Control},
  volume  = {65},
  number  = {5},
  pages   = {2016--2031},
  year    = {2020},
  month   = may,
  doi     = {10.1109/TAC.2019.2926160},
  url     = {https://doi.org/10.1109/TAC.2019.2926160}
}

@techreport{dalke_hufford2005_markov,
  author      = {Dalke, Roger and Hufford, George},
  title       = {Analysis of the Markov Character of a General Rayleigh Fading Channel},
  institution = {National Telecommunications and Information Administration (NTIA), U.S. Department of Commerce},
  year        = {2005},
  number      = {TM-05-423},
  url         = {https://its.ntia.gov/publications/download/TM-05-423.pdf},
  note        = {Technical Memorandum}
}

@inproceedings{pimentel2002its_fsc,
  author    = {Pimentel, Cecilio and Falk, Tiago H. and Lisboa, Luciano},
  title     = {Finite-State Markov Modeling of Flat Fading Channels},
  booktitle = {International Telecommunications Symposium (ITS 2002)},
  address   = {Natal, Brazil},
  year      = {2002},
  doi       = {10.14209/its.2002.576},
  url       = {https://biblioteca.sbrt.org.br/articles/3956}
}

@article{pimentel2004rician_fsmc,
  author  = {Pimentel, Cecilio and Falk, Tiago H. and Lisboa, Luciano},
  title   = {Finite-State Markov Modeling of Correlated Rician-Fading Channels},
  journal = {IEEE Transactions on Vehicular Technology},
  volume  = {53},
  number  = {5},
  pages   = {1491--1501},
  year    = {2004},
  month   = sep,
  doi     = {10.1109/TVT.2004.832413},
  url     = {https://doi.org/10.1109/TVT.2004.832413}
}

@inproceedings{hasslingerhohlfeld2008ge_fit,
  author    = {Ha{\ss}linger, Gerhard and Hohlfeld, Oliver},
  title     = {The Gilbert-Elliott Model for Packet Loss in Real Time Services on the Internet},
  booktitle = {Proceedings of the 14th GI/ITG Conference on Measurement, Modeling and Evaluation of Computer and Communication Systems (MMB 2008)},
  address   = {Dortmund, Germany},
  year      = {2008},
  month     = apr,
  pages     = {269--286},
  url       = {https://www.kom.tu-darmstadt.de/papers/HH08_1034.pdf}
}

@article{nielsen2019multiconnectivity,
  author  = {Nielsen, Jimmy J. and Leyva-Mayorga, Israel and Popovski, Petar},
  title   = {Reliability and Error Burst Length Analysis of Wireless Multi-Connectivity},
  journal = {arXiv preprint},
  year    = {2019},
  url     = {https://arxiv.org/abs/1909.03875},
  note    = {Uses GE fitting on LTE and Wi-Fi measurement traces}
}

@article{vanvleck_middleton1966_clipped,
  author  = {Van Vleck, J. H. and Middleton, David},
  title   = {The Spectrum of Clipped Noise},
  journal = {Proceedings of the IEEE},
  volume  = {54},
  number  = {1},
  pages   = {2--19},
  year    = {1966},
  month   = jan,
  doi     = {10.1109/PROC.1966.4567},
  url     = {https://doi.org/10.1109/PROC.1966.4567}
}

@techreport{bussgang1952,
  author      = {Bussgang, Julian J.},
  title       = {Crosscorrelation Functions of Amplitude-Distorted Gaussian Signals},
  institution = {MIT Research Laboratory of Electronics},
  year        = {1952},
  number      = {Technical Report 216},
  url         = {https://dspace.mit.edu/handle/1721.1/4847}
}

@article{price1958,
  author  = {Price, Robert},
  title   = {A Useful Theorem for Nonlinear Devices Having Gaussian Inputs},
  journal = {IRE Transactions on Information Theory},
  volume  = {4},
  number  = {2},
  pages   = {69--72},
  year    = {1958},
  month   = jun,
  doi     = {10.1109/TIT.1958.1057444},
  url     = {https://doi.org/10.1109/TIT.1958.1057444}
}

@article{baum1957,
  author  = {Baum, R. F.},
  title   = {The Correlation Function of Smoothly Limited Gaussian Noise},
  journal = {IRE Transactions on Information Theory},
  volume  = {3},
  number  = {3},
  pages   = {193--197},
  year    = {1957},
  month   = sep,
  doi     = {10.1109/TIT.1957.1057415},
  url     = {https://doi.org/10.1109/TIT.1957.1057415}
}

@article{gudmundson1991,
  author  = {Gudmundson, Mikael},
  title   = {Correlation Model for Shadow Fading in Mobile Radio Systems},
  journal = {Electronics Letters},
  volume  = {27},
  number  = {23},
  pages   = {2145--2146},
  year    = {1991},
  month   = nov,
  doi     = {10.1049/el:19911328},
  url     = {https://doi.org/10.1049/el:19911328}
}

@book{goldsmith2005,
  author    = {Goldsmith, Andrea},
  title     = {Wireless Communications},
  publisher = {Cambridge University Press},
  address   = {Cambridge, U.K.},
  year      = {2005},
  doi       = {10.1017/CBO9780511841224},
  url       = {https://doi.org/10.1017/CBO9780511841224}
}

@misc{sheppard1900,
  author = {Sheppard, W. F.},
  title  = {On the Calculation of the Double-Integral Expressing Normal Correlation},
  year   = {1900},
  note   = {Transactions of the Cambridge Philosophical Society, vol. 19, pp. 23--66 (often cited as the origin of early bivariate normal correlation integral results).}
}

\end{document}